\documentclass[aps,prl,reprint,groupedaddress]{revtex4-1}

\usepackage{graphicx}
\usepackage{bm}
\usepackage{amsmath,amssymb,esint}
\usepackage[dvips]{color}

\usepackage[linktocpage=true,bookmarks=true,bookmarksnumbered=true,breaklinks=true,pdfpagemode=Fullscreen,pdfstartview=FitBH]{hyperref}

\newcommand{\InGraph}{\includegraphics[scale=0.8]}

\newcommand{\fr}[2]{\mbox{$\frac{\,{#1}\,}{#2}$}}

\def\bge{\begin{equation}}
\def\ede{\end{equation}}
\def\bga{\begin{aligned}}
\def\eda{\end{aligned}}
\newcommand{\beq}{\begin{equation}}
\newcommand{\eeq}{\end{equation}}
\newcommand{\bq}{\begin{equation}}
\newcommand{\eq}{\end{equation}}
\newcommand{\ba}{\begin{array}}
\newcommand{\ea}{\end{array}}
\newcommand{\beqa}{\begin{eqnarray}}
\newcommand{\eeqa}{\end{eqnarray}}
\newcommand{\beqs}{\begin{subequations}}
\newcommand{\eeqs}{\end{subequations}}

\def\nn{\nonumber}

\def\dis{\displaystyle}

\def\ff{\frac}
\def\f{\frac}
\def\({\left(}
\def\){\right)}

\def\deg{\circ}
\def\End{\end{document}}

\def\L{\mathcal{L}}
\def\Oh{O_h^{}}
\def\Z{\mathbb{Z}}
\def\Zs{\mathbb{Z}_2^s}
\def\Zmt{\mathbb{Z}_2^{\mu\tau}}
\def\ZFL{\mathbb{Z}_4^{\ell}}
\def\ZZ{\mathbb{Z}_2^{}}
\def\ZF{\mathbb{Z}_4^{}}
\def\ot{\otimes}
\def\ep{\epsilon}
\def\th{\theta}
\def\ts{\theta_{12}^{}}
\def\ta{\theta_{23}^{}}
\def\tx{\theta_{13}^{}}
\def\d{\delta}
\def\da{\delta_a^{}}
\def\ds{\delta_s^{}}
\def\dx{\delta_x^{}}

\def\mtau{\mu\tau}

\def\mt{\tilde{m}}

\begin{document}


\title{Octahedral Symmetry with Geometrical Breaking:\\
       New Prediction for Neutrino Mixing Angle $\boldsymbol{\theta_{13}}$ and CP Violation}

\author{{\sc Hong-Jian He}\,$^{a,b,c}$ ~and~ {\sc Xun-Jie Xu}\,$^a$}
\affiliation{$^a$Institute of Modern Physics and Center for High Energy Physics,
             Tsinghua University, Beijing 100084, China\\
             $^b$Center for High Energy Physics, Peking University, Beijing 100871, China\\
             $^c$Kavli Institute for Theoretical Physics China, CAS, Beijing 100190, China}


\begin{abstract}
We propose octahedral group $\,\Oh\,$ as the family symmetry of neutrino-lepton sector.
We find that $\,\Oh\,$ contains subgroups $\,\Zmt\otimes\Zs\,$
and $\,\ZFL\,$ for realizing the bimaximal (BM) mixings, $\,\theta_{23}=\theta_{12}=45^\deg$\,
and $\,\theta_{13}=0^\deg\,$,\, 
where $\,\Zmt\otimes\Zs\,$ and $\,\ZFL\,$  serve as the residual symmetries of
neutrinos and charged leptons, respectively. We present geometric interpretations of
BM mixing in the octahedron, and construct natural geometrical breaking of $\,\ZFL\,$,\,
leading to nontrivial deviations from the BM mixings.  Our theory makes truly simple
predictions of a relatively large reactor angle,
$\,\tx \simeq 45^\deg -\ts = 7.5^\deg-13.7^\deg\,$ ($3\sigma$),
the nearly maximal atmospheric angle and the approximate maximal Dirac CP violation.
These agree well with the current neutrino data, and will be further probed
by the on-going and upcoming oscillation experiments.
\\[1.5mm]
Phys.\ Rev.\ D\,(2012), Rapid Communications, in Press [\,{arXiv:1203.2908}\,].
\end{abstract}




\maketitle


\noindent
{\bf 1.~Introduction}
\vspace*{2mm}

The observed three families of neutrinos, leptons and quarks in nature
exhibit very distinctive and puzzling masses and mixing structures, posing
great challenges to particle physics today. Unraveling the underlying family
symmetry holds the best promise to overcome such challenges.
For more than a decade, various oscillation experiments have discovered massiveness of
neutrinos and large atmospheric and solar neutrino mixing angles ($\ta,\,\ts$),
leaving the reactor mixing angle $\,\tx\,$ much smaller and least determined.
Inspecting the current oscillation data\,\cite{fit2012}
shows that the atmospheric angle is around maximal, $\,\ta = O(45^\deg)\,$,\,
and the solar angle $\,\ts\,$ is large, but has significant deviation from its maximal value,
$\,45^\deg - \ts =O(10^\deg)\,$.\, Strikingly, the recent T2K data \cite{T2K}
point to a nonzero $\,\tx\,$ at $2.5\sigma$ level with central value
$\,\tx = 9.7^\deg\,(11.0^\deg)\,$ for normal (inverted) mass ordering,
which is on the same order of the deviation $\,45^\deg - \ts =O(10^\deg)\,$.\,
Subsequently, Minos \cite{Minos} and Double Chooz \cite{DC}
found further indications of a nonzero $\,\tx\,$.\,
The Daya Bay collaboration \cite{DYB2012} newly announced a $5.2\sigma$
discovery of the nonzero $\,\tx\,$,
$\,\sin^22\tx = 0.092\pm 0.016(\text{stat})\pm 0.005(\text{syst})\,$,\,
with central value $\,\tx = 8.8^\deg\,$ and the $5\sigma$ allowed range,
\beqa
\label{eq:DYB-5s}
1.6^\deg ~<~ \tx ~<~ 12.4^\deg \,. 
\eeqa
Shortly afterwards, RENO \cite{DYB2012} found nonzero $\,\tx\,$ at $4.9\sigma$ level,
$\,\sin^22\tx = 0.113\pm 0.013(\text{stat})\pm 0.019(\text{syst})\,$,\,
with central value $\,\tx = 9.8^\deg\,$,\, and allowed $5\sigma$ range,
\beqa
\label{eq:RENO-5s}
0^\deg ~\leqslant~ \tx ~<~ 14.1^\deg \,. 
\eeqa
Both data are consistent with the earlier results of
T2K\,\cite{T2K}, Minos\,\cite{Minos}, and Double Chooz\,\cite{DC} last year.

From the available oscillation data\,\cite{fit2012}-\cite{DYB2012},
we observe a striking pattern of neutrino mixing angles,
\beqa
\ta \,=\, O(45^\deg ), ~~~~
\tx \,=\, O(45^\deg\! - \ts ) \,=\, O(10^\deg)\,,~~~~
\eeqa
where the second relation can be reexpressed in radian,
$\tx = O(\frac{\pi}{4} - \ts ) \,=\, O(0.2)$.\,
Hence, we are strongly motivated to take the
bimaximal (BM) mixing\,\cite{BM}\cite{AF-rev}
as our leading order (LO) structure,
\beqa
\label{eq:LO}
\th_{12}^{(0)} \,=\, \th_{23}^{(0)} \,=\, \frac{\pi}{4} , ~~~~~
\th_{13}^{(0)} \,=\, 0\,,
\eeqa
and then define the small deviations,
$\,\da \equiv \fr{\pi}{4} - \ta\,$,\, $\,\ds \equiv \fr{\pi}{4} - \ts\,$,\,
and $\,\dx \equiv \tx - 0\,$,\,
as the next-to-leading order (NLO) perturbation,
\beqa
\label{eq:NLO}
\da \,\approx\, 0\,,~~~
\dx \,\sim\, \ds  \,=\, O(0.2)\,.
\eeqa

As shown before, in the lepton mass-eigenbasis,
the neutrino sector exhibits a generic LO symmetry $\,\Zmt\otimes\Zs\,$,\,
where $\,\Zmt\,$ dictates $\,(\ta,\,\tx)=(\fr{\pi}{4},\,0)\,$ and
$\,\Zs\,$ determines $\,\ts\,$ by its group parameter $\,k=\tan\ts\,$
(with $k$ being a general real parameter of the 3-dimensional representation
of $\Zs$\,) \cite{He:2011kn}.
We find that the LO mixing pattern (\ref{eq:LO}) is a specific
realization of $\,\Zmt\otimes\Zs\,$ with the group parameter
$\,k=\tan\th_{12}^{(0)}=1\,$.\, We also note that the previously much studied
finite groups \cite{AF-rev}\cite{BM-S4}\cite{BM-S4b},
such as $A_4$ \cite{A4} and $S_4$ \cite{S4}, etc,
do not contain $\,\Zmt\otimes\Zs(k=1)\,$ as the subgroup.
In this work, we demonstrate that the octahedral group $\,\Oh\,$
can unify $\,\Zmt\otimes\Zs(k=1)\,$ and
thus predict the LO structure (\ref{eq:LO}).
We further construct a natural geometrical breaking of the residual
symmetries of $\,\Oh\,$,\, leading to a quantitative prediction of the
NLO structure (\ref{eq:NLO}), as well as the maximal Dirac CP violation.

\vspace*{4mm}
\noindent
{\bf 2.~Geometrical\,Formulation\,of~Bimaximal\,Mixing}
\vspace*{2mm}


Most previous studies focused on the finite groups
$A_4$ \cite{A4} and $S_4$ \cite{S4}, in order to realize the
tri-bimaximal (TBM) mixing ansatz\,\cite{AF-rev}
which is a special realization of the generic product group
$\,\Zmt\otimes\Zs\,$ by choosing the group parameter of $\Zs$ to be,
$\,k=\tan\ts = \fr{1}{\sqrt{2}}\,$ ($\ts\simeq 35.3^\deg$),
as shown in \cite{He:2011kn}.
Another simple realization of $\,\Zmt\otimes\Zs\,$ is,
$\,k=\tan\ts = \fr{2}{3}\,$ ($\ts\simeq 33.7^\deg$)
\cite{He:2011kn}\cite{Ge:2010js},
whose $\ts$ agrees with the current data even better than the TBM value.
But any such choice of $\,k\in\Zs\,$ is independent of $\,\Zmt\,$ (which dictates
$\,\tx = 0\,$ and $\,\ta =\fr{\pi}{4}\,$), and thus does not give the prediction
of $\,\tx\,$.\,
Different from these studies, the present work will start from the
BM mixing (\ref{eq:LO}) as our LO structure, which corresponds to
$\,\Zmt\otimes\Zs(k=1)\,$.\, We find the octahedral group ${\Oh}$
to be a minimal group that contains $\,\Zmt\otimes\Zs(k=1)\,$ as subgroups.

\begin{figure}[t]
\centering
\InGraph{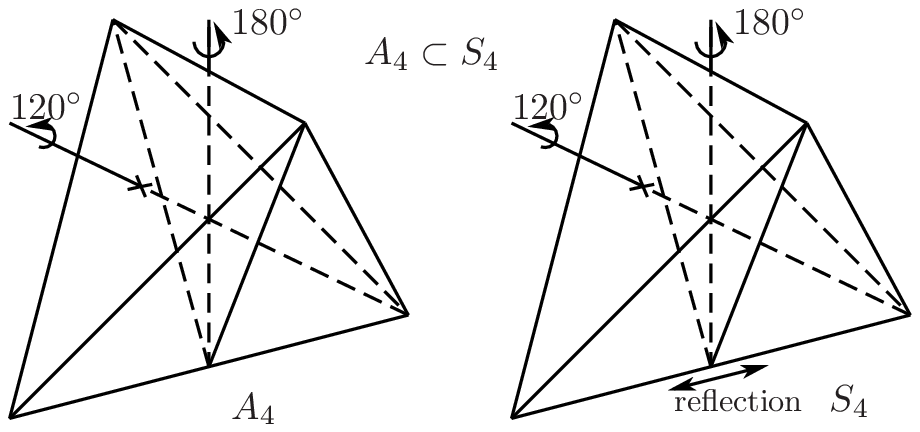}
\\[3mm]
\InGraph{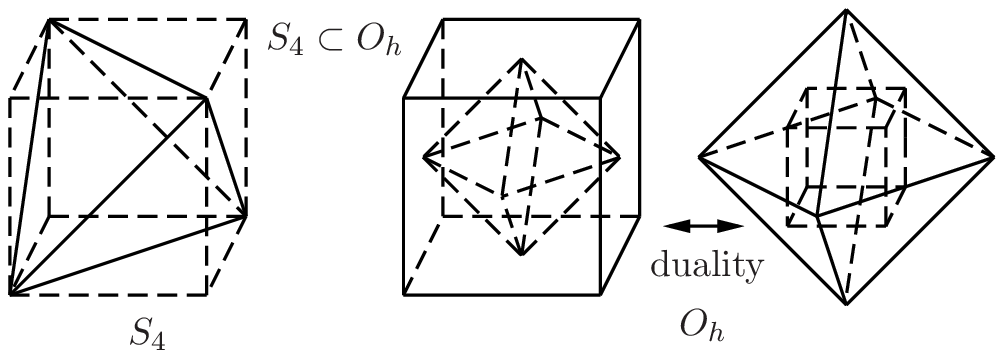}
\caption{Geometric interpretations of $A_4$, $S_4$ and $\Oh$, as well as their relations.
$A_4$ only includes rotations of regular tetrahedron and $S_4$ is the full symmetry group
of regular tetrahedron, thus $\,A_4 \subset S_4\,$.\, Moreover, $S_4$ is a subgroup of
$\Oh$ because the tetrahedron can be contained in a cube, which is the dual form of octahedron.
The duality means a cube can be embedded into a octahedron with its 8 vertices
being the face-centers of the octahedron, and vice versa.}
\label{fig1:a4s4}
\vspace*{-4mm}
\end{figure}

Since $A_{4}$ and $S_{4}$ are much used in the literature\,\cite{AF-rev},
we first analyze the relationships among $\,A_{4}$,\, $S_{4}$\, and $\,\Oh$\,.\,
The $A_{4}$ is the rotational group of a regular tetrahedron, and has
12 elements in total, which includes four $120^{\circ}$ rotations plus
their inverses, and three $180^{\circ}$ rotations.
$S_{4}$ is the full symmetry of a regular tetrahedron and has
24 elements, including the same rotations as $A_4$ and additional reflection.
These are interpreted by the two upper plots in Fig.\,\ref{fig1:a4s4}.
The difference between $A_{4}$ and $S_{4}$
is the additional reflection in $S_4$, which corresponds to $\Zmt$.
Fig.\,\ref{fig1:a4s4} provides an intuitive geometrical explanation
on why $\Zmt$ appears accidentally in the $A_{4}$ models.

A similar situation happens for the BM mixing where the $\Zmt$ appears
accidentally when the BM mixing is obtained from an $S_{4}$ model\,\cite{BM-S4}.
For this we also have a geometric interpretation, as presented by the
lower plots of Fig.\,\ref{fig1:a4s4}. The reason that an $S_{4}$ model may
obtain BM mixing is not due to its tetrahedral property,
but because it is isomorphic to chiral octahedral group $O$,\, a
subgroup of octahedral group $\,\Oh\,$.\, The $\Oh$ is the full
symmetry group of a regular octahedron and isomorphic to $\,O\otimes \Z_2'\,$
or $\,S_4\otimes \Z_2'\,$,
where $\,\Z_2'\,$ is a discrete inversion symmetry.
The group $\Oh$ is well-known in crystal physics.

\begin{figure}[t]
\centering
\includegraphics[scale=1.1]{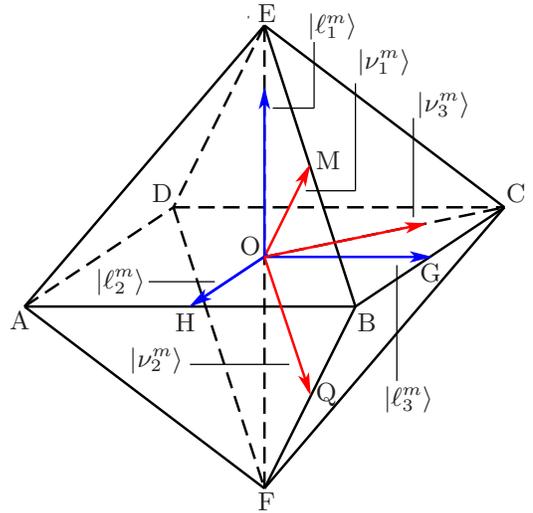}
\caption{Assignments for the left-handed neutrino mass-eigenstates
$(\left|\nu_1^{m}\right>,\,\left|\nu_2^{m}\right>,\,\left|\nu_3^{m}\right>)$\,
and the left-handed charged lepton mass-eigenstates
$(\left|\ell_1^{m}\right>,\,\left|\ell_2^{m}\right>,\,\left|\ell_3^{m}\right>)$
in the regular octahedron, where Q, M, H, and G are the midpoints of the corresponding
edges.}
\label{fig2:nuL-basis}
\vspace*{-5mm}
\end{figure}

Then, we present a geometrical formulation of the BM mixing (\ref{eq:LO})
in the octahedral structure.
Regarding the space in Fig.\,\ref{fig2:nuL-basis} as flavor space,
we assign the left-handed neutrino mass-eigenstates
$(\left|\nu_1^{m}\right>,\,\left|\nu_2^{m}\right>,\,\left|\nu_3^{m}\right>)$
and the left-handed charged lepton mass-eigenstates
$(\left|\ell_1^{m}\right>,\,\left|\ell_2^{m}\right>,\,\left|\ell_3^{m}\right>)$
to certain vectors in the regular octahedron with proper orientations.
Thus, the transition amplitude of $\,\left|\nu_{j}^{m}\right\rangle\,$ being an $\,\left|\ell_i^{m}\right>\,$-flavor
neutrino is $\,\left\langle \ell_i^{m}|\nu_{j}^{m}\right\rangle\,$.\,
We can compute this via inner products of the corresponding geometrical vectors,
\begin{equation}
\hspace*{-3mm}
U_0^{} =\!
\begin{pmatrix}
\left\langle \ell_1^{m}\right|\\[1mm]
\left\langle \ell_2^{m}\right|\\[1mm]
\left\langle \ell_3^{m}\right|
\end{pmatrix} \!\!
\begin{pmatrix}
\left|\nu_{1}^{m}\right\rangle  \!&\! \left|\nu_{2}^{m}\right\rangle
\!&\! \left|\nu_{3}^{m}\right\rangle
\end{pmatrix}
= \!\left(\!
\begin{array}{ccr}
\frac{1}{\sqrt{2}\,} & \frac{-1}{\sqrt{2}\,} & 0~\,
\\[2mm]
\frac{1}{2} & \frac{1}{2} & \frac{-1}{\sqrt{2}\,}
\\[2mm]
\frac{1}{2} & \frac{1}{2} & \frac{1}{\sqrt{2}\,}
\end{array}
\!\right) \!\!,~
\label{eq:U0}
\end{equation}
which is just the leptonic mixing matrix in the charged current,
also called Pontecorvo-Maki-Nakagawa-Sakata (PMNS) matrix \cite{PMNS}.
We readily recognize that (\ref{eq:U0})
just results in the LO structure of BM mixing (\ref{eq:LO}).

It is interesting to note
that tetrahedron, cube and octahedron discussed above all belong to
Platonic solids which are related to the classical elements of fire, earth and air,
respectively, according to the ancient Greek philosopher Plato.
The aesthetic beauty and symmetry of these objects
have been appreciated by human since thousands of years ago.

\vspace*{4mm}
\noindent
{\bf 3.~Application of $\,{\Oh}$ to Neutrinos and Leptons}
\vspace*{3mm}

\begin{figure}[t]
\centering
\includegraphics[scale=1]{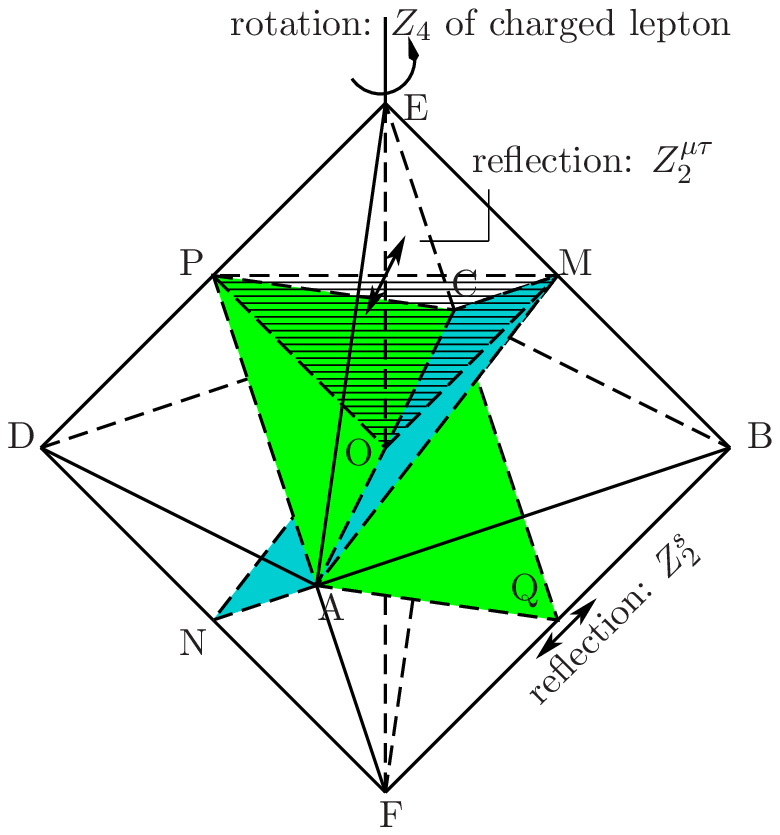}
\caption{Neutrino symmetry $\,\Zmt\ot\Zs$\, and charged
lepton symmetry $\,\ZFL\,$ under octahedral group $\Oh$.
The $\,\Zmt\,$ corresponds to reflections through the plane
PNQM (marked by hachured lines), and the $\,\Zs\,$
generates reflections through the green plane PAQC.
Reflections through the three marked planes (hachured, green, blue)
generate all symmetries in the neutrino sector.
The $\,\ZFL\,$ corresponds to rotations by $90^{\circ}$'s
around the axis $\protect\overrightarrow{\text{FE}}$.}
\label{fig3:nu-symm}
\vspace*{-4mm}
\end{figure}

The full symmetry group of regular octahedron, $\Oh$,
can be generated by two of its elements $\,s\,$ and $\,t\,$
with the \emph{definition relations,}
\beqa
s^{4} = t^{2} = (st)^{6} = (ts^{2}ts)^{2} = 1 \,.\,
\label{eq:st}
\eeqa
The geometrical meanings of $\,s\,$ and $\,t\,$ are $90^{\circ}$
\emph{rotoreflection} about the axis $\overrightarrow{\text{OE}}$
and $180^{\circ}$ rotation about the axis $\overrightarrow{\text{OM}}$,\,
respectively (Fig.\,\ref{fig3:nu-symm}).
Here, by $90^{\circ}$ \emph{rotoreflection}, it means a $90^{\circ}$ rotation
around the axis $\overrightarrow{\text{OE}}$
and then making a reflection about the plane ABCD.

The $O_{h}$ has 48 elements which can be divided into 10 \emph{conjugacy classes}.
Elements in the same class have similar geometrical meanings.
It has 10 irreducible representations (IRs) and the sum of squares of their
dimensions $D_j$ equals the order of $\,\Oh\,$,~
%
$\dis\sum_{j=1}^{10}D_{j}^{2} = 48 \,,$\, 
%
which has the solution, $D=\{1,1,1,1,2,2,3,3,3,3\}$.  
This shows that $\Oh$ has
four 1-dimension (1d) IRs, denoted by $\mathbf{1}$, $\mathbf{1}'$,
$\overline{\mathbf{1}}$, $\overline{\mathbf{1}}'$;\,  two 2d IRs,
denoted by $\mathbf{2}$, $\overline{\mathbf{2}}$;\, and four 3d IRs,
denoted by $\mathbf{3}$, $\mathbf{3}'$, $\overline{\mathbf{3}}$,
$\overline{\mathbf{3}}'$.\,
The IR $\mathbf{3}$ is the fundamental representation, under which
the elements $\,s\,$ and $\,t\,$ take the forms,
\begin{equation}
\label{eq:ST-rep}
S =\!\left(\!\begin{array}{ccc}
-1\\
 &  & -1\\
 & 1
\end{array}\!\right)\!, ~~~~~
T = -\!\left(\!\begin{array}{ccc}
 &  & 1\\
 & 1\\
1
\end{array}\!\right)\!.~~~~~
\end{equation}
Here for notational distinction, we use uppercase (lowercase) letters
for the representations (elements).
The 3d representations (\ref{eq:ST-rep}) can be directly inferred
according to the geometrical meanings of $\,s\,$ and $\,t\,$.\,
In Fig.\,\ref{fig3:nu-symm}, we define $\,\overrightarrow{\text{OE}}$\,
as $x$-axis, $\,\overrightarrow{\text{OA}}$\, as $y$-axis,
and $\,\overrightarrow{\text{OB}}$\, as $z$-axis.

The $\,\Oh\,$ flavor symmetry breaking is expected to occur at a
higher energy scale and leave the residual symmetries in the low energy
mass-matrices of neutrinos and leptons.
For this work, we study the low energy effective theory
and its phenomenological predictions, where the detailed high-scale
dynamics of $\Oh$ breaking is irrelevant.
In our effective theory, the flavor symmetry breaking is explicit
(which is similar to the well-established explicit breakings of discrete
$P$ and $CP$ symmetries in the SM).
Such breaking may arise from the spontaneous symmetry breaking via
proper flavoron fields at high scales\,\cite{AF-rev} or due to certain
unknown new mechanism, but the associated heavy states are all integrated
out from the effective theory and thus irrelevant to our low energy study.
Since spontaneous breaking of discrete flavor symmetry still suffers other
problems and complications including the well-known domain wall problem,
we prefer explicit breaking for our effective theory.
In our construction, the residual symmetry for leptons is
$\,\ZFL$\,,\, and is generated by the $90^{\circ}$ rotation $\,r_{\ell}^{}$\, around
the axis $\overrightarrow{\text{OE}}$ in Fig.\,\ref{fig3:nu-symm}.
For neutrinos, the residual symmetry is $\,\Zmt\ot\Zs\,$,\, and is generated
by the two reflections, $\,r^{}_{\mtau}\,$ and $\,r^{}_{s}\,$,\,
as depicted in Fig.\,\ref{fig3:nu-symm}.
We can express them as below,
\begin{equation}
\begin{cases}
~\ZFL \!: & r_{\ell}^{} ~=~ t(st)^{2} \,,
\\[1mm]
~\Zmt \!\!:~~ & r^{}_{\mtau} ~=~ sts^{3}tst \,,~~~~~
\\[1mm]
~\Zs \!: & r^{}_{s} ~=~ sts^{3}ts \,.
\end{cases}
\label{eq:Z4-Z2mt-Z2s}
\end{equation}
Then, we derive their 3-dimensional representations,
\beqa
\hspace*{-10mm} &&
R_{\ell}^{}\,=\!\left(\!\begin{array}{ccr}
1~ & 0 & 0\\
0~ & 0 & -1\\
0~ & 1 & 0
\end{array}\!\right)\!,~~~~
R_{s}=\!\left(\!\!\begin{array}{rrr}
0~ & 0 & -1\\
0~ & 1 & 0\\
-1~ & 0 & 0
\end{array}\!\right)\!,
\nn\\[1.5mm]
\hspace*{-10mm} &&
R^{}_{\mtau} =\, \text{diag}(1,-1,1)\,.
\label{eq:rep-Z4ZmtZs}
\eeqa

In our construction, we assign the three generations of leptons
to the triplet representation $\,\mathbf{3}$\,.\,
Then, we can write down the effective Lagrangian
for lepton and neutrino mass-terms at the leading order,
\beqa
\mathcal{L}_0^{} & = &
\mu_{1}^{}\ell_{1}^{c}\ell_{1}^{}+\mu_{2}^{}(\ell_{2}^{c}\ell_{2}^{}
+\ell_{3}^{c}\ell_{3}^{}) +\mu_{3}^{}(\ell_{2}^{c}\ell_{3}^{}
-\ell_{3}^{c}\ell_{2}^{})
\label{eq:L0mass-lep-nu}
\\ 
&& +m_{11}^{}(\nu_{1}^{}\nu_{1}^{} \!+ \nu_{3}^{}\nu_{3}^{})
   +m_{22}^{}\nu_{2}^{}\nu_{2}^{}
   +2m_{13}^{}\nu_{1}^{}\nu_{3}^{} + \text{h.c.},~~
\nn
\eeqa
where \,$(\nu_{1}^{},\nu_{2}^{},\nu_{3}^{})^{T}\equiv\nu$\, and
\,$(\ell_{1}^{},\,\ell_{2}^{},\,\ell_{3}^{})^{T}\equiv \ell$\,
are left-handed neutrinos and leptons,
and $(\ell_{1}^{c},\,\ell_{2}^{c},\,\ell_{3}^{c})\equiv \ell^{c}$ denote
the right-handed anti-leptons.
The Lagrangian (\ref{eq:L0mass-lep-nu}) is invariant under the above
residual symmetries, i.e., the $\,\ZFL\,$ for leptons,
\begin{equation}
\ell ~\to~ R_{\ell}^{}\,\ell\,,\, ~~~~~
\ell^{c} ~\to~ \ell^{c}R_\ell^{\dagger}\,,
\label{eq:Z4-transf}
\end{equation}
and $\,\Zmt\ot\Zs$\, for Majorana neutrinos,
\begin{equation}
\label{eq:Zmt-Zs-transf}
\nu\to R_{\mtau}^{}\nu \,, ~~~\textrm{or},~~~
\nu\to R_s^{}\nu \,.
\end{equation}
The mass matrices of leptons and neutrinos from (\ref{eq:L0mass-lep-nu})
can be diagonalized \cite{note1} by the mapping matrices
\,$U_{\ell 0}^{}$\, and \,$U_{\nu 0}$,\, respectively,
\beqa
\hspace*{-3mm}
U_{\ell 0}^{} = \!\left(\!\begin{array}{ccc}
1 & 0 & 0\\[1mm]
0 & \frac{i}{\sqrt{2}} & \frac{-i}{\sqrt{2}}\\[1.5mm]
0 & \frac{1}{\sqrt{2}} & \frac{1}{\sqrt{2}}
\end{array}\!\right) \!\!,~~~
U_{\nu 0}^{} = \!\left(\!\begin{array}{ccc}
\frac{1}{\sqrt{2}} & \frac{-1}{\sqrt{2}} & 0
\\[1.5mm]
0 & 0 & 1
\\[1.2mm]
\frac{1}{\sqrt{2}} & \frac{1}{\sqrt{2}} & 0
\end{array}\!\right)\!\widetilde{U}_0'\,,~~~~~~~
\label{eq:UL-Unu}
\eeqa
where $\widetilde{U}_0'$ is a diagonal Majorana-phase matrix absorbing possible
phases from the neutrino mass-eigenvalues.
Thus, we can deduce the PMNS mixing matrix,
\beqa
V_0^{} ~=~ U^{\dag}_{\ell 0}U_{\nu 0}^{} ~=~ U_0^{}U_0' \,,
\label{eq:V0-LO}
\eeqa
where $\,U_{0}^{}\,$ exactly agrees to the BM mixing (\ref{eq:U0})
(as derived earlier from the geometrical construction
in Fig.\,\ref{fig2:nuL-basis}), and
$\,U'_0=\text{diag}(1,\,1,\,i)\widetilde{U}_0'$\,
is the Majorana-phase matrix.

We also note that in the lepton-mass-eigenbasis, the symmetry transformation
matrices of (\ref{eq:rep-Z4ZmtZs}) become,
$\,(R_\ell',\,R_{\mu\tau}',\,R_s')
 = U^{\dag}_{\ell 0}(R_\ell^{},\,R_{\mu\tau}^{},\,R_s^{})U_{\ell 0}^{}$\,,\, with
%
\beqa
\hspace*{-5mm} &&
R_{\ell}' \,=\, \textrm{diag}(1,\,i,\,-i)\,,
\label{eq:Rlf}
\nn\\[1mm]
\hspace*{-7mm} &&
R_{\mu\tau}' =
\left(\begin{array}{ccc}
1 & 0 & 0 \\
0 & 0 & 1 \\
0 & 1 & 0
\end{array}\right)  \!, ~~~~~
R_{s}' =
\left(\!\begin{array}{rrr}
0~ & \frac{-1}{\sqrt{2}} & \frac{-1}{\sqrt{2}}
\\[1.5mm]
\frac{-1}{\sqrt{2}} & \f{1}{2} & -\f{1}{2}
\\[1.5mm]
\frac{-1}{\sqrt{2}} & -\f{1}{2} & \f{1}{2}
\end{array}\right) \!.~~~
\label{eq:Rl-Rmt-Rs}
\eeqa
%
Here, the transformation $R_{\mu\tau}'$ is just the conventional form
of $\Z_2^{\mu\tau}$ in the lepton-mass-eigenbasis and corresponds to
the symmetry under the $\nu_\mu^{}$ and $\nu_\tau^{}$ exchange.

As a final note of illustration, we give a type-I seesaw realization
of the neutrino masses in (\ref{eq:L0mass-lep-nu}), where we assign the
Higgs doublet $\Phi$ of the standard model (SM) as singlet of
the $O_h^{}$ group, and the three heavy right-handed neutrinos
$(\nu_{1}^{c},\nu_{2}^{c},\nu_{3}^{c})\equiv\nu^{c}$
as triplet, $\nu^{c}\rightarrow\nu^{c}R_{\mu\tau,s}^{T}$\,
under $\,\Zmt\ot\Zs$\ transformation. Thus, we can write down the seesaw
Lagrangian to replace the second row of (\ref{eq:L0mass-lep-nu}),
\begin{eqnarray}
\mathcal{L}_{\textrm{ss}}^{}\! & = &
      \mt_{11}^{}(\nu_{1}^{c}\nu_{1}^{} \!+\! \nu_{3}^{c}\nu_{3}^{})
\!+\! \mt_{22}^{}\nu_{2}^{c}\nu_{2}^{}
\!+\! \mt_{13}^{}(\nu_{1}^{c}\nu_{3}^{}\!+\!\nu_{3}^{c}\nu_{1}^{})~~~
\nonumber\\[1.5mm]
 & &
 - M_{11}(\nu_{1}^{c}\nu_{1}^{c}\!+\!\nu_{3}^{c}\nu_{3}^{c})
 \!-\! M_{22}\nu_{2}^{c}\nu_{2}^{c}
 \!-\! 2M_{13}\nu_{1}^{c}\nu_{3}^{c} \,,
\label{eq:seesaw}
\end{eqnarray}
where the Dirac masses arise from the products of the
corresponding Yukawa couplings and Higgs vacuum expectation
value $\left<\Phi\right>$.\,
Integrating out the heavy $\nu_j^{c}$\,'s from (\ref{eq:seesaw}), we
derive the light neutrino seesaw-mass-terms in the effective Lagrangian
(\ref{eq:L0mass-lep-nu}) with the coefficients,
%
\beqa
m_{11}^{} &=&
\frac{\,(\mt_{11}^2\!\!+\!\mt_{13}^{2})M_{11}^{}\!- 2\mt_{11}^{}\mt_{13}^{}M_{13}^{}}
     {M_{11}^{2}\!-\!M_{13}^{2}} \,,
\nn
\\[1mm]
m_{22}^{} &=& \frac{\mt_{22}^{2}}{M_{22}^{}} \,,
\label{eq:seesaw-m}
\\[1mm]
m_{13}^{} &=&
\frac{\,-(\mt_{11}^2\!\!+\! \mt_{13}^{2})M_{13}^{}\!\!+\! 2\mt_{11}^{}\mt_{13}^{}M_{11}^{}\,}
     {M_{11}^{2}\!-\!M_{13}^{2}} \,.
\nn
\eeqa
%
At low energies our effective theory has the same particle content
as the SM with light massive neutrinos. The heavy states associated with
possible flavor breakings are integrated out around the seesaw scale,
and thus cause no visible flavor changing effect at low energies.

\vspace*{4mm}
\noindent
{\bf 4.~Geometrical\,Breaking\,and\,Physical\,Predictions}
\vspace*{3mm}

Sec.\,3 demonstrated how the LO BM mixing (\ref{eq:LO})
is predicted by the residual symmetries $\,\Zmt\ot\Zs\,$ and $\,\ZFL\,$
of the octahedral group $\,\Oh\,$.\,
In this section, we will construct a natural geometrical breaking of
$\,\ZFL\,$ in the lepton sector, and derive physical predictions.

Inspecting the residual symmetries $\,\Zmt\ot\Zs\,$ and $\,\ZFL\,$,\, we see that
there are two fundamental ways to introduce the breaking: one is to break the
$\,\ZFL\,$ in the lepton sector, and another is to break $\,\Zmt\ot\Zs\,$
in the neutrino sector (as in \cite{He:2011kn}\cite{Ge:2010js}).
In either way of the breaking, it is generally recognized that charged leptons
always have the maximal residual symmetries
$\,U(1)\otimes U(1)\otimes U(1)\supset \ZF$\,,\, and neutrinos always
have the maximal residual symmetries
$\,\ZZ\otimes \ZZ \otimes \ZZ \supset \Z_2^{a}\otimes \Z_2^{b}\,$,\,
even though after any possible breaking
the $\,\ZF\,$ will differ from the original
$\,\ZFL\,$ and $\,\Z_{2}^{a}\otimes \Z_{2}^{b}\,$
is no longer the $\,\Zmt\ot\Zs$\,.\,
The breaking of the original $\,\ZFL\,$
or $\,\Zmt\ot\Zs$\, always accompanies the emergence
of a new $\,\Z_4'\,$ or $\,\Z_2'\ot\Z_2'\,$.\,
Hence, adding a residual symmetry breaking term $\,\d\L\,$ to the LO
Lagrangian $\,\L_0^{}\,$ just means changing the corresponding residual symmetry
from $\,\ZFL\,$ to another $\,\Z_4'\,$, or from
$\,\Zmt\ot\Zs$\, to another $\,\Z_2'\ot\Z_2'\,$.\,
It usually turns out that the emergent new residual symmetries in the new Lagrangian
$\,\L_0^{}+\d\L\,$,\,  such as $\,\Z_4'\,$ or $\,\Z_2'\ot\Z_2'\,$,\,
may be no longer physically useful, but the modification
of the original symmetry predictions based on the LO Lagrangian $\,\L_0\,$ due
to the residual symmetry breaking term $\,\d\L\,$ does lead to physically testable
corrections.

\begin{figure}[t]
\centering
\includegraphics[scale=1.1]{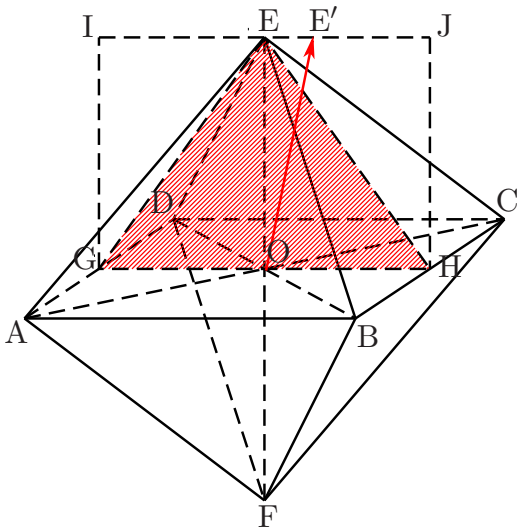}
\caption{Geometric breaking of $\,\ZFL\,$ is realized by pulling the
$\ZFL$ axis $\protect\overrightarrow{\text{OE}}$ to
$\protect\overrightarrow{\text{OE}'}$,
where the point E$'$ is in the line IJ.}
\label{fig:GBreak}
\vspace*{-4mm}
\end{figure}

For our study, we find that the simplest way comes from
a natural geometrical breaking of $\,\ZFL\,$ in lepton sector.
This geometrical breaking is constructed by simply rotating the axis
$\,\overrightarrow{\text{OE}}$\, of $\,\ZFL\,$
to another axis $\,\overrightarrow{\text{OE}}'$\,
associated with a new $\,\Z_4'\,$,\, where we define the
small rotation angle $\,\angle\text{EOE}'\equiv \sqrt{2}\ep\,$ with a characteristic
breaking parameter $\,\ep\,$.\, We illustrate this geometrical breaking in
Fig.\,\ref{fig:GBreak}, where the axis $\overrightarrow{\text{OE}}$ rotates to
$\overrightarrow{\text{OE}}'$ within the plane GIJH
and G (H) is the middle point of the line AD (BC).
Thus we can derive the $\ZFL$ breaking Lagrangian,
\beqa
\label{eq:dL-GB}
\delta\mathcal{L} &\,=\,&
-\ep\left[(\mu_{1}^{}\!-\!\mu_{2}^{}\!-\!\mu_{3}^{})
\ell_{1}^{c}\ell_{2}^{}
+(\mu_{2}\!-\!\mu_{1}\!-\!\mu_{3})\ell_{1}^{c}\ell_{3}^{} \right.
\\
&&\left.
+(\mu_{1}^{}\!-\!\mu_{2}^{}\!+\!\mu_{3}^{})\ell_{2}^{c}\ell_{1}^{}
+(\mu_{2}^{}\!-\!\mu_{1}^{}+\mu_{3}^{})\ell_{3}^{c}\ell_{1}^{}\right]
+ \text{h.c.},~~~~
\nn
\eeqa
where $\,\mu_{1,2,3}^{}\,$ have appeared in the LO Lagrangian
(\ref{eq:L0mass-lep-nu}).  We find that the small breaking parameter $\ep$ indeed
generates the NLO deviation via
$\,\ep\simeq 45^{\circ}-\ts =O(0.2)\,$.\,
We have verified that $\,\mathcal{L}_0+\delta\mathcal{L}\,$
is invariant under transformations of $\,\widetilde{R}_\ell^{}\in\Z_4'\,$.\,
Geometrically, $\,\widetilde{R}_\ell^{}\,$ is a rotation around
the axis $\overrightarrow{\text{OE}}'$ by $90^{\circ}$,
and we deduce up to $O(\ep )$,
\begin{equation}
\widetilde{R}_{\ell}^{} ~=\left(\!\begin{array}{ccr}
~1 & -2\epsilon & 0
\\ 
~0 & 0 & -1
\\ 
2\epsilon & 1 & 0
\end{array}\right)  \!.
\label{eq:cp31}
\end{equation}

Combining the breaking term (\ref{eq:dL-GB}) with the LO Lagrangian
(\ref{eq:L0mass-lep-nu}), we derive the full PMNS mixing matrix up to the NLO,
\begin{equation}
V \simeq
\left(\!\begin{array}{ccc}
\frac{1+\epsilon}{\sqrt{2}} & \frac{-1+\epsilon}{\sqrt{2}} & i\epsilon
\\[1.5mm]
\frac{1-\epsilon}{2}\!-\!\frac{i\epsilon}{2}~
& \frac{1+\epsilon}{2}\!+\!\frac{i\epsilon}{2}
& -\frac{1}{\sqrt{2}}
\\[1.5mm]
\frac{1-\epsilon}{2}\!+\!\frac{i\epsilon}{2}~
& \frac{1+\epsilon}{2}\!-\!\frac{i\epsilon}{2}
& ~\frac{1}{\sqrt{2}}
\end{array}\!\right)\! U' ,~~~~~
\label{eq:V-NLO}
\end{equation}
where $\,U'=U_0'$\, as given below (\ref{eq:V0-LO})
since the symmetry breaking Lagrangian
(\ref{eq:dL-GB}) does not affect the neutrino mass matrix.
It is clear that $\,V\,$ differs from the LO result (\ref{eq:V0-LO})
by $\,O(\ep)\,$ corrections.
We can also break the $\,\ZFL$\, by rotating
$\overrightarrow{\text{OE}}$ in the direction perpendicular to the plane
GIJH, but the difference can be absorbed by flipping the sign of
Dirac CP phase. Finally, from (\ref{eq:V-NLO}), we
derive the following solution, up to $O(\ep)$ corrections,
\beqa
\label{eq:predict-1}
\da \,\simeq\, 0\,, ~~~~
\ds \,\simeq\, \dx \,\simeq\, \ep\,, ~~~~
\d_D^{} \,\simeq\, \pm\ff{\pi}{2} \,,~~~~~~~~~~~
\eeqa
where as we defined before, the three NLO parameters
$\,(\da,\,\ds,\,\dx) \equiv \(\ff{\pi}{4} - \ta,\, \ff{\pi}{4} - \ts,\, \tx\)\,$,\,
and $\,\d_D^{}\,$ is the Dirac CP phase angle.
From (\ref{eq:predict-1}), we infer the truly simple and beautiful
predictions below,
\beqa
\label{eq:predict-2}
\tx \,\simeq\, 45^\deg\! - \ts\,, ~~~~
\ta \,\simeq\, 45^\deg , ~~~~
\d_D^{} \,\simeq\, \pm 90^\deg .~~~~~~
\eeqa
The first relation above strikingly predicts the crucial reactor angle
$\,\tx\,$ in terms of the precisely measured solar
angle $\,\ts\,$,\, without explicit dependence on the breaking parameter
$\,\ep\,$ itself.
The current global fit\,\cite{fit2012} gives the
$3\sigma$ range of solar angle,
$\,31.3^{\circ} < \ts < 37.5^{\circ}\,$,\,
with a central value $\,\ts =34.5^\deg\,$.\,
With this, our first relation of (\ref{eq:predict-2}) predicts the
$3\sigma$ range of the reactor angle,
\beqa
7.5^\deg ~<~ \tx ~<~ 13.7^\deg \,.
\eeqa
This is rather distinct and agrees well with the new
data from Daya Bay and RENO reactor experiments \cite{DYB2012}.

\begin{figure}[t]
\centering
\includegraphics[width=8.3cm,height=6.4cm,clip=true]{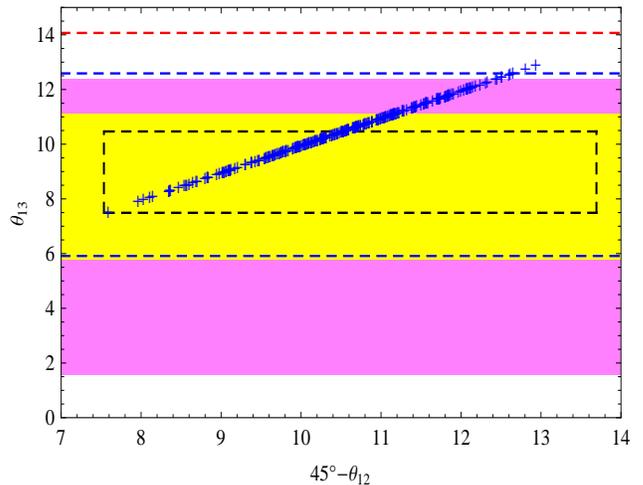}
\vspace*{-1mm}
\caption{Prediction of $\,\tx\,$ as a function of $\,45^\deg\!-\ts\,$,\,
with 300 samples (marked by blue $+$).
The yellow (pink) region corresponds to the $3\sigma$ ($5\sigma$) new limit of
Daya Bay\,\cite{DYB2012}. The $3\sigma$ ($5\sigma$) new limit of RENO\,\cite{DYB2012}
is given by the blue (red) dashed lines.
The black dashed box depicts the allowed $3\sigma$ ranges
by the global fit\,\cite{fit2012}.}
\label{fig:t13-t12}
\vspace*{-4mm}
\end{figure}

We further plot the predicted $\,\tx\,$
as a function of $\,45^\deg\! - \ts\,$ in Fig.\,\ref{fig:t13-t12},
where we have input $\,\ts\,$ from the global fit\,\cite{fit2012}
by generating 300 random samples with Gaussian distribution.
Fig.\,\ref{fig:t13-t12} also shows the $3\sigma$ constraint
by the global fit\,\cite{fit2012} on the allowed parameter space, as marked
by black dashed box. We further present the $3\sigma$ and $5\sigma$
limits of $\,\tx\,$ from the new Daya Bay data\,\cite{DYB2012},
as given by the horizontal yellow and pink bands, respectively.
The $3\sigma$ ($5\sigma$) limits of RENO data\,\cite{DYB2012}
are shown by the horizontal blue (red) dashed lines.
Our prediction on $\,\tx\,$ (marked by blue $+$) lies in
the upper half of the allowed regions by Daya Bay and RENO data,
and will be further tested by the upcoming data of Daya Bay\,\cite{DYB2012},
T2K\,\cite{T2K}, Double Chooz\,\cite{DC} and RENO\,\cite{RENO} experiments.

\begin{figure}[t]
\centering
\includegraphics[width=8cm,height=6cm,clip=true]{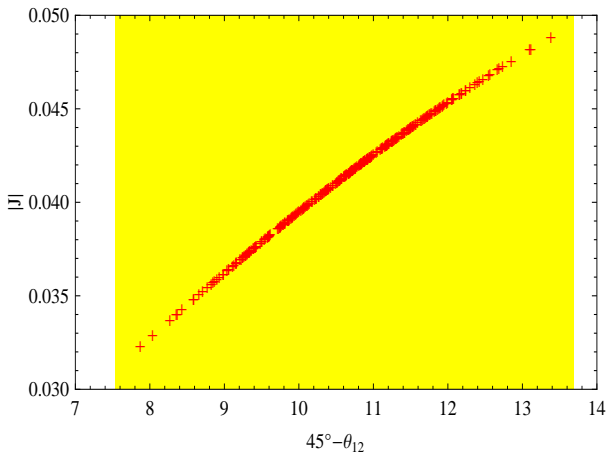}
\vspace*{-1mm}
\caption{Prediction of Jarlskog invariant $J$ as a function of $\,45^\deg\!-\ts\,$,\,
 with 300 samples (marked by red $+$). The yellow band represents the $3\sigma$ allowed
 ranges of $\,45^\deg\!-\ts\,$ from the global fit\,\cite{fit2012}.}
\label{fig:Jarlskog}
\end{figure}

We also note that under the seesaw generation of
light neutrino masses via (\ref{eq:seesaw})-(\ref{eq:seesaw-m}),
it is natural to have the $O_h$ breaking and $\ZFL$ breaking realized
at or above the seesaw scale.
Then, we can further include the small effects of renormalization group (RG) running
for mixing angles (\ref{eq:predict-2}). We find that the RG running effects are possible 
to reduce $\ta$ and $\,\ts +\tx\,$ by about $2^\deg$ at low energy. Thus, we have
$\,\ta \simeq 43^\deg$\, and $\,\tx \simeq 43^\deg -\ts$\,,\,
which will agree better with the recent global fit\,\cite{fit2012}.

Finally, we further compute the Jarlskog invariant $\,J\,$  \cite{J}
as a measure of leptonic CP violations. With (\ref{eq:predict-2}),
we derive $\,J\,$ in terms of the solar mixing angle $\,\ts\,$ alone,
\beqa
J &\,=\,& \fr{1}{8}\sin 2\ta\sin 2\ts \sin 2\tx \cos\tx \sin\d_D^{}
\nn\\[1.5mm]
  & \,\simeq\, & \pm\fr{1}{16}\sin 4\ts \,.
\label{eq:J-1}
\eeqa
From the global fit \cite{fit2012},
we derive the $3\sigma$ range
of the predicted Jarlskog invariant,
\begin{equation}
\left|J\right| \,=\, 0.031-0.050\,.
\label{eq:J-2}
\end{equation}
With Eq.\,(\ref{eq:J-1}) and inputting the global fit\cite{fit2012} of $\,\ts\,$,
we plot our prediction of $\,J\,$ as a function of $\,45^\deg\! -\ts\,$ 
in Fig.\,\ref{fig:Jarlskog}, where the yellow region shows the $3\sigma$ limits
on $\,45^\deg \!-\ts\,$ from the global fit.
Our distinctive prediction of the nearly maximal Dirac CP violation will be further
probed by the upcoming long baseline oscillation experiments.

\vspace*{5mm}
\noindent
{\bf 5.~Conclusions}
\vspace*{3mm}

Understanding flavor symmetries behind three families of the
standard model fermions in nature holds the best promise
to unravel their mysterious masses and mixing structures.
In this work, we proposed octahedral group $\Oh$ as the flavor symmetry
of neutrino-lepton sector, whose residual symmetries $\,\Zmt\otimes\Zs\,$ and $\,\ZFL\,$
prescribe the neutrinos and charged leptons, respectively.

\vspace*{1mm}

The current neutrino data strongly motivate us to consider the bimaximal mixing (BM)
of (\ref{eq:LO}) as a good LO structure and the deviations from the BM mixing
in Eq.\,(\ref{eq:NLO}) can nicely serve as the NLO perturbation.
In Sec.\,2-3, we presented the geometric interpretations of $O_{h}^{}$ and its
residual symmetries (Fig.\,\ref{fig1:a4s4} and Fig.\,\ref{fig3:nu-symm}).
We found that the unique octahedral group $\Oh$ naturally includes
the subgroups $\,\Zmt\otimes\Zs\,$ and $\,\ZFL\,$ for generating the BM mixing
structure (\ref{eq:U0}) or (\ref{eq:V0-LO}).

\vspace*{1mm}

In Sec.\,4, we further constructed a natural geometrical breaking of
$\,\ZFL\,$ as in Fig.\,\ref{fig:GBreak} and Eq.\,(\ref{eq:dL-GB}).
The resultant PMNS mixing matrix $V$ was derived in Eq.\,(\ref{eq:V-NLO}).
Our theory makes truly simple new predictions of a relatively large reactor angle,
$\,\tx \simeq 45^\deg -\ts = 7.5^\deg-13.7^\deg\,$ ($3\sigma$),
the nearly maximal atmospheric angle,
and the approximate maximal Dirac CP violation.
These are shown in Eq.\,(\ref{eq:predict-2}) and presented in
Fig.\,\ref{fig:t13-t12}. Especially, our new prediction of the reactor angle
$\,\tx\,$ agrees well with the new Daya Bay discovery (\ref{eq:DYB-5s}) and
the RENO limit (\ref{eq:RENO-5s}).  
Finally, we predicted Jarlskog invariant $\,J\,$  in terms of a single
mixing angle $\,\ts\,$ via Eq.\,(\ref{eq:J-1}). With this we derived its $3\sigma$
range, $\,|J|=0.031-0.050\,$,\, as in Eq.\,(\ref{eq:J-2}).
The correlation between Jarlskog invariant $\,J\,$ and the deviation
$\,45^\deg\!-\ta\,$ is depicted in Fig.\,\ref{fig:Jarlskog}.
All our new predictions agree well with the current neutrino data,
and will be further tested by the on-going and upcoming oscillation experiments.

\vspace*{1mm}

\vspace*{2mm}
\noindent
{\bf Acknowledgments}
\\[2mm]
We thank C.\ S.\ Lam for discussing the octahedral group
during his visit to Tsinghua HEP Center in the fall of 2011.
We are grateful to Vernon Barger and Werner Rodejohann for discussing this subject.
We also thank Luca Merlo and Davide Meloni for discussing Refs.\,\cite{BM-S4}\cite{BM-S4b}.
This work was supported by National NSF of China (under grants 11275101, 11135003,
10625522, 10635030) and National Basic Research Program (under grant 2010CB833000),


\begin{thebibliography}{99}
\bibitem{note1}
We note that diagonalizing the leading-order lepton mass-matrix in
(\ref{eq:L0mass-lep-nu}) gives mass-eigenvalues,
$m_e^{}=\mu_1^{}$, and $\,m_{\mu,\tau}^{} = |\mu_2 \pm i\mu_3|$\,.\
Here the electron has no mixing with $(\mu,\tau)$ leptons and its mass-parameter
$\mu_1^{}$ is fixed by data.
The parameters $\mu_{2,3}^{}$ are complex in general.
For a simple illustration, choose
$\mu_2^{}$ to be real and $\,\mu_3^{} = i\bar{\mu}_3^{}\,$ to be imaginary.
Thus, the $(\mu,\tau)$ masses take the simple form,
$\,m_{\mu,\tau}^{} = |\mu_2 \mp \bar{\mu}_3|$\,.\
Given the ratio of experimental $(\mu,\tau)$ mass-values
$\,m_{\tau}^{}/m_{\mu}^{}\simeq 16.8 = O(10)\,$,\, we find
that this is realized by inputting
$\,\bar{\mu}_3^{}/\mu_2^{}\simeq 0.89$, or $1.13$,
which is naturally $O(1)$.
Including the NLO lepton mass-terms in (\ref{eq:dL-GB}),
we explicitly verified that
the above LO mass-eigenvalues remain unchanged.
Hence our construction can accommodate the observed lepton masses.
\end{thebibliography}

\end{document}